\documentclass[conference]{IEEEtran}
\usepackage{amsmath,amssymb,amsfonts}
\usepackage{algorithmic}
\usepackage{float}
\usepackage{graphicx}
\usepackage{textcomp}
\usepackage{stfloats}
\usepackage{caption}

\usepackage{xcolor}
\def\BibTeX{{\rm B\kern-.05em{\sc i\kern-.025em b}\kern-.08em
    T\kern-.1667em\lower.7ex\hbox{E}\kern-.125emX}}
    
\usepackage[sorting=none]{biblatex} 
\begin{document}

\title{Feature-Rich Long-term Bitcoin Trading Assistant\\
}

% \author{\IEEEauthorblockN{1\textsuperscript{st} Jatin Nainani}
% \and
% \IEEEauthorblockN{2\textsuperscript{nd} Nirman Taterh}
% \and
% \IEEEauthorblockN{3\textsuperscript{rd} Md. Ausaf Rashid}

% }
\author{%
Jatin Nainani\textsuperscript{1}\textsuperscript{*}, Nirman Taterh\textsuperscript{1}\textsuperscript{*}, 
Md. Ausaf Rashid\textsuperscript{1}, Ankit Khivasara\textsuperscript{1}\textsuperscript{a}\\[1ex]
\textsuperscript{1}K. J. Somaiya College of Engineering \\
\textsuperscript{a}Corresponding Author \\
\textsuperscript{*}These authors have contributed equally to this work
}

\maketitle

\begin{abstract}
% Trading is an essential aspect of commerce in human history. % In this project we discover the plausibility of the use of RL in an algorithmic trade system.
% Firstly, the daily closing prices and volume of bitcoin in the market are used.
% To evaluate the current value of the prices,
For a long time predicting, studying and analyzing financial indices has been of major interest for the financial community. Recently, there has been a growing interest in the Deep-Learning community to make use of reinforcement learning which has surpassed many of the previous benchmarks in a lot of fields. Our method provides a feature rich environment for the reinforcement learning agent to work on. The aim is to provide long term profits to the user so, we took into consideration the most reliable technical indicators. We have also developed a custom indicator which would provide better insights of the Bitcoin market to the user. The Bitcoin market follows the emotions and sentiments of the traders, so another element of our trading environment is the overall daily Sentiment Score of the market on Twitter. The agent is tested for a period of 685 days which also included the volatile period of Covid-19. It has been capable of providing reliable recommendations which give an average profit of about 69\%. Finally, the agent is also capable of suggesting the optimal actions to the user through a website. Users on the website can also access the visualizations of the indicators to help fortify their decisions.
\end{abstract}


\section{Introduction}
Cryptocurrency is a virtual currency which is secured by cryptography which uses cryptographic functions to facilitate financial transactions and form a system to store \& transfer value. It leverages blockchain technology in order to be decentralized. Its most important feature is that it's not controlled by a central authority like the government or bank to uphold or maintain it. Bitcoin, which was released in 2009 is the first decentralized cryptocurrency.

\noindent While there are numerous solutions and trading bots that try to generate profits by trading and through short term patterns, there is a need for an investing assistant which tries to maximise profits by considering long term trends and real-time user sentiments.

\noindent It helps give the investor a perspective of the overall market situation and accordingly enables them to plan their investment and allocate an appropriate percentage of their financial portfolio to cryptocurrencies and Bitcoin.

\noindent It is important to note that more than 98\% percentage of short-term traders do not make a profit net of fees. This is because it is complicated to time the market and make the right trading decision every time. This is because there are too many variables involved in predicting the short-term price of any asset, more so in the case of Bitcoin since it is one of the most volatile financial assets in the market.

\noindent The critical difference between a trading bot and an investing assistant is:  In the case of the trading bot, the final trade decision is made by the bot itself, while in the investing assistant, the final investment decision is on the investor. 

There are numerous challenges involved in building a robust and accurate reinforcement learning model that's a good enough representation of a cryptocurrency market's environment. Cryptocurrency price actions are extremely volatile and depend on a large amount of real-world and statistical factors. We will start by outlining the overall flow and architecture of the model. We then move forward to discuss the implementation of various features like Twitter sentiments, technical indicators and the custom index. Finally, all these features will be utilized in the reinforcement learning model. The model will be evaluated using robust evaluation metrics like entropy loss, policy loss and value loss. The final product is hosted on a website for the user to get the recommendations and gain insights from the visualizations.
% \subsection{Brief description of project undertaken}
% In this project, we have two objectives primarily. The first one is to help the investors visualize the current market with the help of various price indicators vis, RSI, EMA, SMA and Custom Index based on the bull market support band. 

% \noindent The second part involves generating a buy, sell or hold recommendation after taking into account the aforementioned price indicators, the current market position of the investor and the overall sentiment of the users and the investor. 

% \noindent We aim to take a balanced overview of the various price indicators, user sentiments and investor's market position, to generate an optimized recommendation that will maximize risk-adjusted returns over the long term.

% \section*{Organization of the report}
% This report is structured as follows. Chapter 2 introduces the features and problems of current solutions in the field. In Chapter 3, we analyze our objectives and explore the fine details of Reinforcement Learning and Sentiment Analysis. In Chapter 4, we delve into the implementation steps of our system, descripting the data and features used in the process. Finally, in Chapter 5, we present our conclusions.

\section{Literature Survey}

Behavioural economists like Daniel Kahneman and Amos Tversky have established that decisions, including financial consequences, are impacted by the emotions and not just value alone \cite{kahneman2013prospect} . Dolan et al.'s work in ”Emotion, Cognition, and Behavior” further supports that emotions highly impact decision-making \cite{dolan2002emotion} . Insights from these researchers open up prospects to find advantages through various tools like sentiment analysis as it shows that demand for goods, and hence their price, may be impacted by more than its economic fundamentals. Panger et al. found that the Twitter sentiment also correlated with people’s overall emotional state \cite{panger2017emotion} . 

\noindent Few other researchers have studied the effectiveness of sentiment analysis of tweets. Pak et al. showed that separating tweets into either positive or negative or neutral categories will result in effective sentiment analysis  \cite{pak2010twitter} . Having established that emotions influence decisions, that social media can impact decisions, and that sentiment analysis of social media can accurately reflect the larger population’s opinions towards something. Dennis et al. collected valence scores on tweets about companies within the S\&P 500 and found that the scores correlated with stock prices \cite{sul2014trading} . De Jong et al. analyzed the minute-by-minute stock price as well as the tweet data for the 30 stocks within the DOW Jones Industrial Average and found that 87\% of stock returns were influenced by the tweets \cite{de2017returns} .

\noindent Similar work has been done on cryptocurrencies after their introduction to see if such methods effectively predict cryptocurrency price changes. In the paper by Stenqvist et al., the authors describe how they collected tweets related to Bitcoin and Bitcoin prices from May 11 to June 11 in 2017. First, tweets were cleaned of non-alphanumeric symbols (using “\#” and “@” as examples of symbols removed). Then the tweets which were not relevant or were determined to be very influential were removed from the analysis. The authors then used VADER (the Valence Aware Dictionary and Sentiment Reasoner) to finally analyze the sentiment of each tweet and hence classify it as negative, neutral, or positive. Only tweets that could be considered positive or negative were kept in the final analysis \cite{stenqvist2017predicting}. The sentiment analysis done in this project builds off everything above but is unique, and we solve the problem of giving the tweets a weight.

\noindent Isaac et al. \cite{madan2015automated} used 10 second and 10-minute BTC price data from OKcoin and Coinbase API to accurately predict the bitcoin prices for short term periods. They focused on 25 features relating to the Bitcoin price and payment network over the course of five years. With this, training was done for Binomial GLM, SVM and Random Forests and these were compared with each other through metrics like sensitivity, specificity, accuracy and precision. The random forest model was able to beat all the models in all metrics except precision, where Binomial GLM bested it.

\noindent Zhengyao et al. \cite{jiang2017deep} used a trading period of 30 minutes. Their primary focus was on portfolio management rather than price prediction, and the trading experiments were done in the exchange Poloniex. The Reinforcement Learning framework used a deterministic policy gradient algorithm. Agent was the software portfolio manager performing trading actions in a financial market environment. The reward function aimed to maximize the average logarithmic cumulative return R. For policy functions, three different methods were tested. CNN, RNN and LSTM. Performance metrics were: Accumulative portfolio value, Sharpe ratio. However with a  fairly high commission rate of around 0.25\% in the backtests, the framework was able to achieve at least 4-fold returns in 50 days.

\noindent Otabek et al. \cite{sattarov2020recommending} created a recommendation system for cryptocurrency trading with Deep Reinforcement Learning. Data was taken from cryptodatadownload and focused on three currencies Bitcoin (BTC), Litecoin (LTC), and Ethereum (ETH). The environment was responsible for accounting for stock assets, money management, model monitoring, buying and holding or selling stocks and finally calculating the reward for actions taken. While the agent ran every day, constantly observing the environment to then choose an action with the policies learnt during the training. The environment monitoring, determining actions with policies, computing the gradient, recording and calculating rewards with discounted rewards,and updating the systems' networks with gradients are all jobs that the agent is responsible for. The experiment on Bitcoin via DRL application shows that the investor got 14.4\% net profits within one month. Similarly, tests when carried out on Ethereum and Litecoin also finished with 74\% and 41\% profit, respectively.

\noindent Fengrui et al. \cite{liu2021bitcoin} regarded the transaction process as actions.The returns were regarded as awards and prices are considered to be states to align with the idea of reinforcement learning. A Deep Reinforcement Learning Algorithm - Proximal Policy Optimization (PPO) was used for high-frequency bitcoin trading. They first compared the results between advanced machine learning algorithms like support vector machine (SVM), multi-layer perceptron (MLP), long short-term memory (LSTM), Transformers and temporal convolutional network (TCN), by applying them to real-time bitcoin price and the experimental results demonstrated that LSTM outperforms. It was then decided to use LSTM as the policy for the PPO algorithm. The approach was able to trade the bitcoins in simulated environment paired with synchronous data and obtained a 31.67\% more return than that of the best benchmark, improving the benchmark by 12.75\%.

\noindent Gabriel et al. \cite{borrageiro2022recurrent} created an agent that learns to trade the XBTUSD (Bitcoin versus US Dollars) perpetual swap derivatives contract on BitMEX on an intraday basis. The cryptocurrency agent realized a total return of 350\%, net transaction costs over five years, 71\% of which is down to funding profit. The annualized information ratio that it achieves is 1.46. The echo state network provides a scalable and robust feature space representation. \\
\noindent Joonbum et al. \cite{leem2020action} developed an action-specialized expert model designed specifically for each reinforcement learning action: buy, hold, and sell. Models were constructed by examining and defining different reward values that correlate with every action under some specific conditions, and the investment behaviour is reflected with each expert model. To verify the performance of this technique, the profits of the proposed system are compared to the profits of the common ensemble and single trading systems. In addition, sensitivity was checked with three varying reward functions: profits and Sharpe ratio and the Sortino ratio. All the experiments had been conducted with Hang Seng Index, S\&P500 and Eurostoxx50 data. The model was 39.1\% and 21.6\% more effective than the single and common ensemble models, respectively. 
\noindent We can conclude a few points from the above literature survey. There is a distinct lack of a consistent environment, leading to some really restrictive while others are too free and ill-defined. Most of the agents are restricted to a single type of market. The variety of preprocessing techniques used led to the question of whether the improvement in the metric was the result of the model or the data fed. Most models are treated as a complete black box with a lot of hyperparameter tuning. Perhaps some form of explainable AI might find some use here to convince the investors and help them understand on what basis our model recommends actions.

\section{Approach} 

\bigbreak
\subsection{Block Diagram}
The block diagram is shown in Figure 1. It describes the flow of the project.
\begin{figure*}
    \centering
    \includegraphics[width=\textwidth,height=8cm]{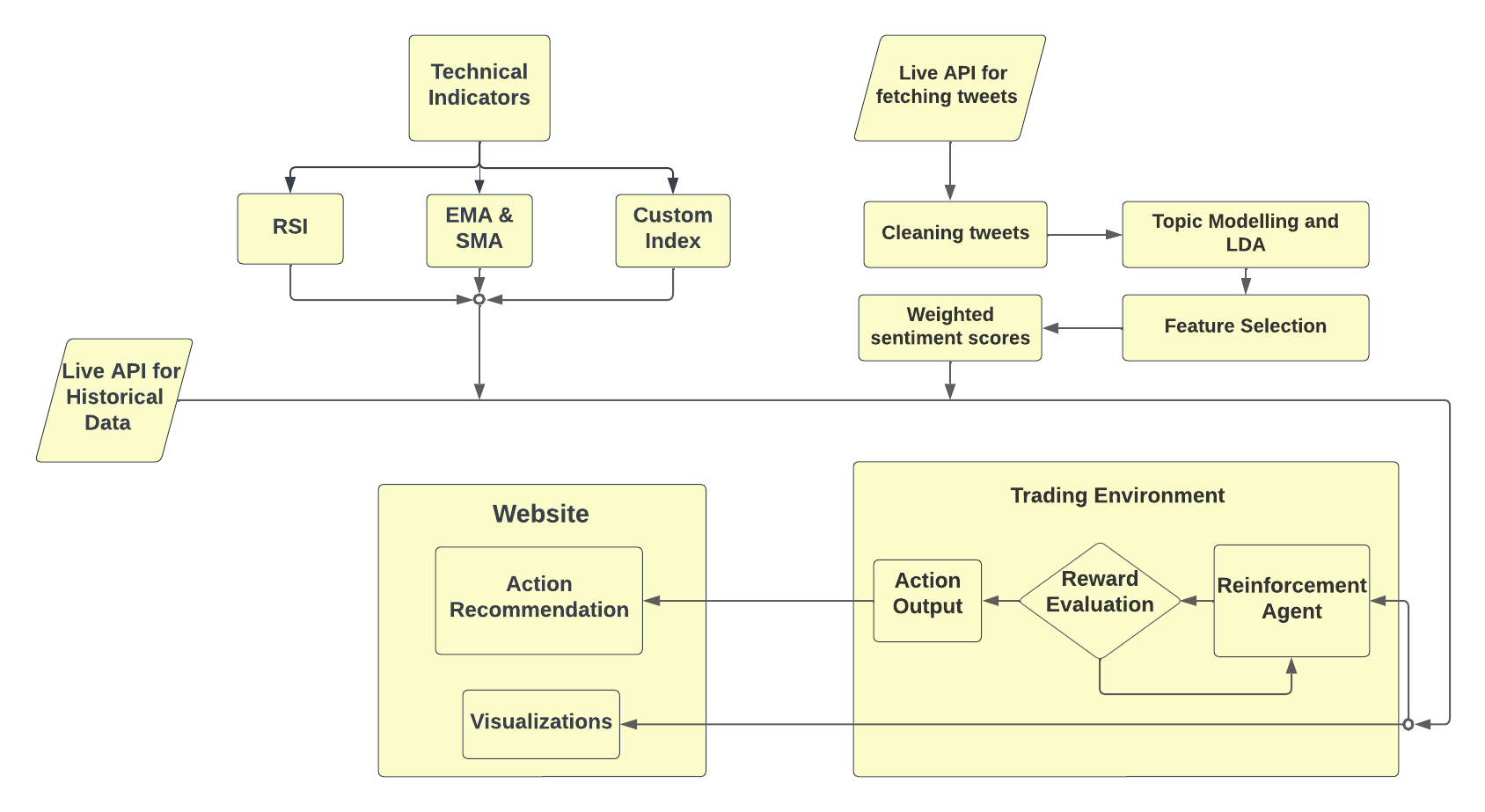}
    \label{fig:galaxy}
    \caption[Flow Chart]{\textit{Block Diagram}}
\end{figure*}
\bigbreak
\subsection{Data Description}
\smallbreak
\subsubsection{Bitcoin Historical Data}

\noindent We obtained the data on past prices from Cryptocompare as it regularly reviews exchanges, carefully monitoring the market to deliver the most accurate pricing indices. Bitcoin prices are known to be volatile, and the cryptocurrency market is still evolving. The main difference between BTC prices and usual stock prices is that the BTC prices change on a much larger scale than the local currencies.

\noindent The overall data collection period is from January 2016 to April 2021 on a daily fidelity which gives us a total of 1918 values. This dataset consists of six different attributes, namely, Date, High, Low, Open, Close, and Volume. High and Low signifies the highest and lowest price at which Bitcoin was traded for that day. Close signifies the last recorded traded price of Bitcoin for that day. Volume here refers to the Bitcoin trading volume, which indicates how many Bitcoins are being bought and sold on specific exchanges on the day. Rows like High and Low are irrelevant for our analysis and can be hence dropped later.

\smallbreak
\subsubsection{Bitcoin related Tweets}

\noindent Twitter is a minefield of information due to the volume and variety of users. This has resulted in Crypto Twitter being significantly influential on cryptocurrency trading decisions. Thus, we collected tweets to perform sentiment analysis on. We used the Twitter API to query tweets by applying to the developer role. However, the rate limit is limited to 10 calls per second, so fetching tweets for a period of 2016/1 to 2021/04 was time-consuming. 

\noindent‘bitcoin,’ ‘BTC,’ ‘\#Bitcoin,’ and ‘\#BTC’ were used as the keyword filters for fetching tweets. Also, all tweets were fetched in the English language. This gave us a total of  4,265,266 tweets to work with. The data contains many columns, including the date of the tweet, username, tweet location, tweet ID, number of replies, number of retweets, number of favorites, the text of the tweet, list of mentions, hashtags, and permalinks. Some of the columns are irrelevant to our analysis and will be dropped later.

\subsection{Technical Analysis}

\smallbreak
\subsubsection{Trend Indicators}

\paragraph{Simple Moving Average}
% \noindent \textbf{Simple Moving Average: }
% \newline
\noindent SMA is just the mean price over the specified period. The mean average is called "moving" because it is plotted on the chart bar-wise, which forms a line that moves along the chart with respect to the average value. If the simple moving average points up, this means that Bitcoin's price is increasing. Down pointing indicates a decrease in the Bitcoin's price.Also, a longer time frame for moving average gives a much smoother simple moving average. A moving average with a shorter-term is more volatile, but the reading is much closer to the source data.

\noindent The shorter the time span(n) used to create the average, the more sensitive it will be to price changes. The longer the time span(n), the less sensitive the average will be to the price changes because more lag is introduced between the SMA and the Bitcoin prices. When it comes to our project, we aim at maximizing profits in the long run. So, we took a longer time span of 21 weeks to calculate our SMA to make sure that it is not sensitive to all the little changes in price but also does not overlook the major price changes. A time span of 21 weeks gives us a very balanced value of SMA, which is well suited for our project.
% \vspace{-0.55cm}
\begin{equation}
    SMA = \frac{(A_1+ A_2+...+A_n)}{n}
\end{equation}
\begin{center}
    $A_n$ is the price of Bitcoin at period $n$ \\
    $n$ is the number of total periods
\end{center} 
% \newline 
 
% \begin{figure}[H]

%     \centering
%     \includegraphics[width=6cm,height=2cm]{images/sma_formula.PNG}
%     \label{fig:galaxy}
%     \caption*{\textit{Formula for SMA}}
% \end{figure}
% % \begin{center}
%     \textit{Formula for SMA}
% \end{center}

\paragraph{Exponential Moving Average}
\noindent Exponential Moving Average is comparable to SMA, gauging trend direction over a time period. On the other hand, SMA calculates average of price data whereas the EMA applies more weight to current data. Hence the EMA follows prices more closely than corresponding SMA for the same.

\noindent EMA is used to determine trend direction and to trade in that direction. When the EMA rises buying when prices dip near the EMA is a good move. When the EMA falls, selling when prices rally towards EMA is a good choice.\\
The formula for EMA at period n is:
\begin{equation}
\begin{aligned}
    EMA_n = &\left(A_n*\left(\frac{S}{1+n}\right)\right) \newline \\ &+ EMA_{n-1} * \left(1-\left(\frac{S}{1+n}\right)\right)
\end{aligned}
\end{equation}
% \vspace{-0.5cm}
\begin{center}
% \begin{figure}[H]
%     \centering
%     \includegraphics[width=9cm]{images/ema_formula.png}
%     \label{fig:galaxy}
%     % \caption[Formula for EMA]{\textit{Formula for EMA}}
% \end{figure}
% \begin{center}
%     \textit{Formula for EMA}
% \end{center}

% \vspace{-1cm}
where, $A_n$ is the price of Bitcoin at period $n$ \\ 
$S$ is the smoothing factor\\
    $n$ is the number of total periods

\end{center}
\noindent Although there are many options to choose from when considering the smoothing factor, we set up a value of 2. A value of 2 gives more credibility to recent data points available. The further we increase the smoothing factor value, the greater influence more recent data has on the moving average. So on testing for different values, we found the value of 2 to give just enough credibility to recent data. The EMA has a shorter delay than the SMA within the same period. Hence we decided to go for a period of 20 weeks for EMA calculations.

\paragraph{Relative Strength Index}
\noindent The RSI is a momentum oscillator that measures the change of price movements. The RSI oscillates between zero and 100. Traditionally the RSI is considered to be overbought when its value is above 70 and oversold when the value is below 30. RSI is also helpful to identify the general trend.

\noindent Value of RSI is calculated with a two steps that starts with the below formula:

\begin{equation}
    RS = \frac{Avg. Gain}{Avg. Loss}
\end{equation}

\begin{equation}
    RSI = 100 - \frac{100}{1 + RS}
\end{equation}
%  \begin{figure}[H]
%     \centering
%     \includegraphics[width=4cm, height=2cm]{images/rsi_formula.png}
%     \label{fig:galaxy}
%     % \caption[Formula for RSI]{Formula for RSI}
% \end{figure}

% \begin{center}
%     \textit{Formula for RSI}
% \end{center}

\noindent The average percentage of the gain or loss during the look-back period is the average gain or loss used in the computation. The formula uses an overall positive value for the average loss. The periods of losses in price are counted as 0 in the computations of average gain, and the periods of increase in price are counted as 0 during the calculation of average loss.

\noindent Usually 14 periods are used to calculate the initial RSI value. Once 14 periods of data is available, the second part of the RSI formula is to be done which smooths the results.

\smallbreak

\subsubsection{Custom Indicator - BMSB}
\noindent BMSB index is a Bollinger band that helps us understand the current valuation of bitcoin against USD, vis., undervalued, overvalued, or fair valued. It ranges from -100 to 100, with the negative side corresponding to extreme undervaluation and the positive side indicating extreme overvaluation with respect to recent price movements.

\noindent Bull Market Support band is a type of Bollinger band that consists of 20w SMA and 21w EMA.

\noindent Usually, when the price of Bitcoin falls below BMSB, there is a bearish momentum in the market, and when the price of Bitcoin holds support or stays above the BMSB, it indicates the onset of a bull market. That being said, being over-extension (+70 to +100 range) does not necessarily imply a bull run, and under-valuation (-70 to -100 range) does not necessarily imply a bearish movement. In fact, it can indicate the end of a bull run and bear market, respectively. Therefore, in our project, when the BMSB index tends to -100, we show it as a buying opportunity, and when it tends to +100, we show it as a time to take profits. 

\paragraph{Algorithm}
Let us introduce a few of the crucial variables that will enable us to formulate the algorithm for the BMSB Index: We define Combined Average ($\mu$) as the average of the current 21 weeks EMA and 20 weeks SMA. We define $K_p$ as the Price Coefficient. The price coefficient decides the extent of the price range from which the Bitcoin price is considered to be normal. The normal price range varies from $\mu(1-K_p)$ to $\mu(1+K_p)$. 
% If the price of bitcoin is within the normal range, we calculate the BMSB Index as: 

% \vspace{1.3cm}
\par\noindent\rule{8cm}{0.4pt} \\
\textbf{Algorithm 1:} BMSB Indicator
\par\noindent\rule{8cm}{0.4pt} \\

\noindent \textbf{Inputs:} Price of Bitcoin (p), Exponential Moving Average (EMA), Simple Moving Average (SMA), Price Coefficient $ (K_P) $ and Scaling Coefficient $ (K_S)$
    
\noindent \textbf{Outputs:} BMSB Index (I)
\newline
   
\noindent \textbf{START} \\
    
    Combined Average ($\mu$) = Average(SMA, EMA)\\
    \indent Upper Bound ($B_U$) = $\mu(1-K_P)$ \\
    \indent Lower Bound ($B_L$) = $\mu(1+K_P)$ \\
    
    \textbf{IF} $ \hspace{0.2cm}p>B_L\hspace{0.4cm} \&\& \hspace{0.4cm}p<B_U\hspace{0.2cm} $ \textbf{then} 
    \vspace{-0.2cm}
\begin{flalign*}
    \hspace{1.1cm}\textrm{I} = \left(\frac{p-\mu}{\mu}\right)\hspace{0.1cm}\frac{K_S}{K_P} \times 100 &&
\end{flalign*}

    \vspace{0.2cm}
    \textbf{ELSE IF} $ \hspace{0.2cm}p<B_L \hspace{0.2cm} $  \textbf{then} 
    \vspace{-0.2cm}
\begin{flalign*}
    \hspace{1.1cm}\textrm{I} = \left[\frac{p(1-K_S)}{\mu(1-K_P)} - 1\right]\times 100 &&
\end{flalign*}

    \vspace{0.2cm}
    \textbf{ELSE} $ \hspace{0.2cm}p>B_U\hspace{0.2cm} $  
    \vspace{-0.2cm}
\begin{flalign*}
    \hspace{1.1cm}\textrm{I} = \left[1-\frac{\mu\hspace{1mm}(1+K_P)\hspace{2mm}(1-K_S)}{p}\right]\times 100 &&
\end{flalign*}

\textbf{end IF} \\
\vspace{0.3cm} \\
\noindent \textbf{STOP} \\

\par\noindent\rule{8cm}{0.4pt} \\

%      \hspace{2cm} \textbf{IF,}
% \begin{equation*}
%     [\hspace{0.2cm}p>\mu(1-K_P)\hspace{0.4cm} \&\& \hspace{0.4cm}p<\mu(1+K_P)\hspace{0.2cm}]
% \end{equation*}

% \hspace{2cm} \textbf{THEN,}
% \begin{equation*}
%     x = \left(\frac{p-\mu}{\mu}\right)\hspace{0.1cm}\frac{K_S}{K_P} \times 100
% \end{equation*}

%     \hspace{2cm} \textbf{ELSE IF,}
% \begin{equation*}
%     [\hspace{0.2cm}p<\mu(1-K_P)\hspace{0.2cm}]
% \end{equation*}

% \hspace{2cm} \textbf{THEN,}
% \begin{equation*}
%     x = \left[\frac{p(1-K_S)}{\mu(1-K_P)} - 1\right]\times 100
% \end{equation*}

%      \hspace{2cm} \textbf{ELSE,}
% \begin{equation*}
%     [\hspace{0.2cm}p>\mu(1+K_P)\hspace{0.2cm}]
% \end{equation*}

% \hspace{2cm} \textbf{THEN,}
% \begin{equation*}
%     x = \left[1-\frac{\mu\hspace{1mm}(1+K_P)\hspace{2mm}(1-K_S)}{p}\right]\times 100
% \end{equation*}

\noindent $K_s$ is defined as the scaling coefficient. The scaling coefficient decides the extent on the BMSB Index scale for the normal price range. It varies from 100(0-$K_s$) to 100(0+$K_s$) and corresponds to the $\mu(1-K_p)$ to $\mu(1+K_p)$ price range. We calculate the BMSB Index as shown in Algorithm 1.

\bigbreak
% \bigbreak
\subsection{Sentiment Analysis}

\subsubsection{Preprocessing}

\smallbreak
\paragraph{Cleaning}
After scraping the tweets, we had to drop some columns that we deemed irrelevant for the sentiment analysis. We kept the columns that we deemed useful, which include the date, username, number of replies, retweets, favorites, and the text of the tweet. Before we are able to start doing any form of sentiment analysis, the tweets collected have to be cleaned.

\noindent Sample tweet before cleaning:
\begin{center}
“Bitcoin Climbs Above 11,316.7 Level, Up 6\% https://yhoo.it/2YV6wKZ \#bitcoin \#crypto  \#blockchain \#btc \#news \#cryptocurrency pic.twitter.com/mPv3rd4kLn \#eth \#ltc \#xrp”
\end{center}

These tweets contain a large amount of noise. Using regex expressions, all these noises were removed. Preprocessing is a very important aspect of sentiment analysis; if we were to pass these raw tweets to our analyzer, chances are it will perform very poorly and take up much more time
 
\noindent Sample Tweet after cleaning:
\begin{center}
    
“Bitcoin Climbs Above Level Up” \\
\end{center}

\noindent Next, we set all the characters to lower cases and also removed stopwords in our tweets. Stop words are commonly used words such as “a,” “the,” and “as,” which provide no valuable information in Sentiment analysis. We made use of the Natural Language Toolkit(NLTK) to get rid of them. This was enough for VADER as it was specially tuned for social media content. 

\paragraph{Topic Modeling and LDA for Ads and Spam handling}
In our first attempt on sentiment analysis, our sentiment analysis did not yield good results as it gave advertisement tweets a high positive sentiment score 
\begin{center}
    E.g. “Free bitcoin faucet with extra bonuses Get free bitcoins now” : +0.8807                             (Using Vader)
\end{center} 
This is an issue as our prediction model was unable to differentiate between a useful tweet and ads. Hence, we identified and tagged these ads by using Latent Dirichlet Allocation (LDA) to cluster and identify possible tweet topics. This is useful as it allows us to gain insight into the type of Bitcoin topics people discuss on Twitter and, within these topics, to identify an ad topic cluster. The reason for choosing LDA is that it is most effective in generalizing to new documents easily. 

\noindent To train our LDA model, the following parameters were varied:
\begin{itemize}
    \item Corpus: Bag of Words generated from the pre-processed tweets
    \item id2word: Dictionary mapping of word ID to words in tweets 
    \item Number of topics: The number of topic clusters to be generated.
\end{itemize}
The model parameters were varied by using different numbers of tweets and the number of topic clusters. Each tweet is presented as a distribution over topics, and every topic is represented as a distribution over words. To evaluate our model parameters, we used the following:
\begin{itemize}

    \item{Visual analysis of clusters with pyLDAvis: }

Through the visual analysis of 10, 6,3,20 topic clusters, we found that 20 topic clusters gave the best result. In the case of 3 topic clusters, none of the clusters overlap in the Intertopic Distance Map indicating the topics were too broad and vague to use them to identify an ad cluster. The histogram on the right represents the most probable topic word for that cluster of tweets. Hence, we increased the cluster size until we found that our clusters gave us unique, most probable topic clusters. As seen here in Fig 2, we can see that the various possible Bitcoin topics are: Bitcoin, market, price, BTC, mining, analyst, new, day, transaction, news, time, analysis, blockchain, dont, eth, payment, exchange, million, social media, resistance. Based on the cluster topics, we suspected that the unusual ‘don’t’ cluster(Fig 3) (cluster 14 on pyLDAvis, topic 13) is likely to be the ad cluster.

\begin{figure}[H]
    \centering
    \includegraphics[width=8cm]{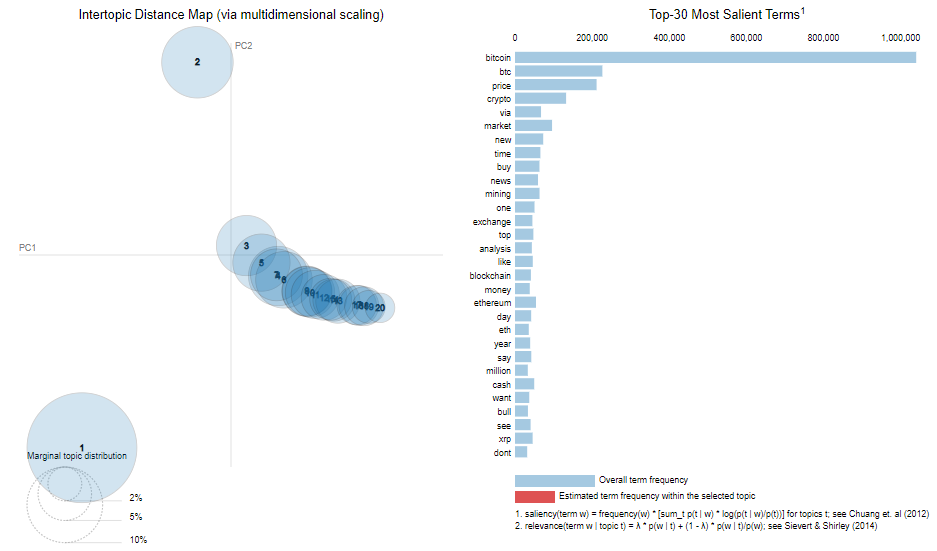}
    \label{fig:galaxy}
    \caption[pyLDA Visualization for 20 Clusters]{\textit{pyLDAvis interactive visualization of Topics generated from LDA. Model parameters: \# of tweets used =All, \# of clusters = 20.}}
\end{figure}

\begin{figure}[H]
    \centering
    \includegraphics[width=8cm]{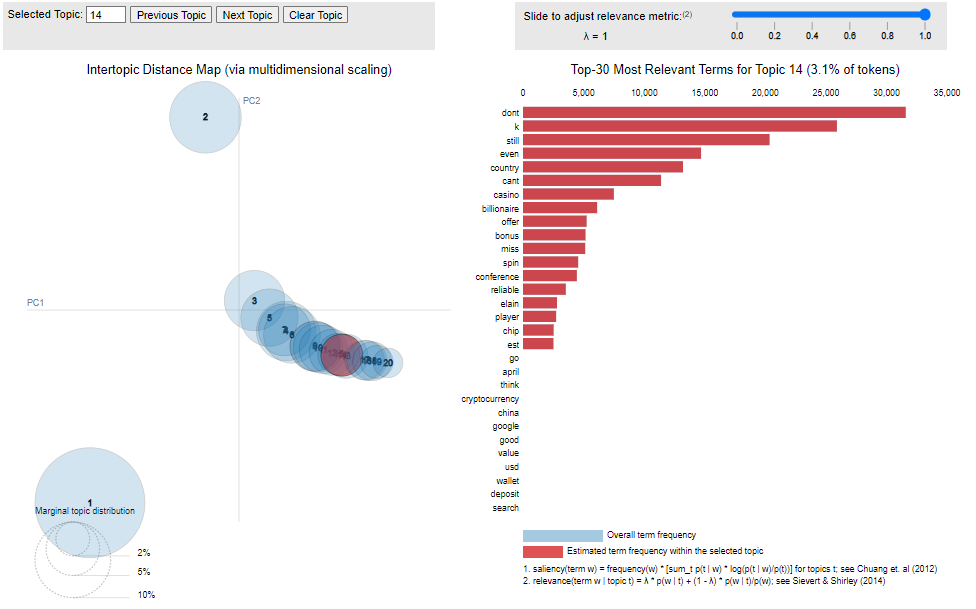}
    \label{fig:galaxy}
    \caption[pyLDA Visualization for 18 Clusters]{\textit{pyLDAvis for suspected ad topic cluster}}
\end{figure}

 \item{Running new unseen documents through our model} 

\end{itemize}
Some examples of Ad tweets include, 'is airdropping den to participants sign up here'; 'claim daily bitcoins at for free'; 'bitcoin black v bitcoin get free coins  value '; 'get free bitcoin'; 'claim free bitcoin now link moon bitcoin is a bitcoin'
 
To confirm our hypothesis, we used tweets that we know are ads (above) and ran it through the LDA model. We wanted to find the model parameters that will result in all 5 tweets’ LDA results returning topic 13 is the most probable topic. Hence, we are able to establish that topic 13 is our ad cluster and our ideal model parameters are to use all the tweets with 20 topic clusters. To tag the ad tweets, we used all tweets and ran it through our LDA model with 20 clusters, and filtered out all Topic 13 tokenized tweets and extracted the words to form an “ad word list” which is used in the preprocessing stage to tag the ad tweets.\\ 

% \begin{figure}
%     \centering
%     \includegraphics[width=7cm]{images/lda1.png}
%     \label{fig:galaxy}
%     \caption[pyLDA Visualization for 3 Clusters]{\textit{pyLDAvis interactive visualization of Topics generated from LDA. Model parameters: \# of tweets used =All, \# of clusters = 3. On the left, is a PCA Plot, where the size of area of the circle refers to the
% importance of each topic over the entire corpus, distance refers to the similarity. On the right, the histogram shows the top 30 most probable topic terms.}
% }
% \end{figure}

\subsubsection{Sentiment Analysis}
 We then compare between VADER and TextBlob, which are lexicon-based approaches for sentiment analysis. We selected a few tweets to compare if VADER or TextBlob performs better. We noticed that the sentiment score for advertisements and spam tweets was mostly quite positive. Hence, we filtered out such tweets, which has been described in detail in the Topic Handling section. From the tests we conducted on the selected few tweets,  VADER works better with things like slang, emojis, etc., which is quite prevalent on Twitter, and it is also more sensitive than TextBlob. So, we choose to go with VADER.

After the sentiment of each Tweet is computed, we take the daily average sentiment, but taking the average sentiment results in the loss of information. For example, a prominent user’s positive tweet (score of +1.0) may have greater influence over public opinion as compared to a common users’ negative tweet (score of 0). To address this issue, we decided to use weighted Sentiment analysis scores.

Feature Selection and Weighted Sentiment Score: 
Our collection of features consists of Tweets text, Time of Tweet, Number of Retweets, Number of Replies, Number of Favorites, Advertisement score, Tweet Volume, and Sentiment Analysis Score (VADER). To prevent the loss of information by taking the average sentiment, we have generated weights to create weighted sentiment scores. Each tweet will have a weighted score. Table 1 shows the Weight Rules we have established.
\begin{table*}
    \centering
    \includegraphics[width=11cm]{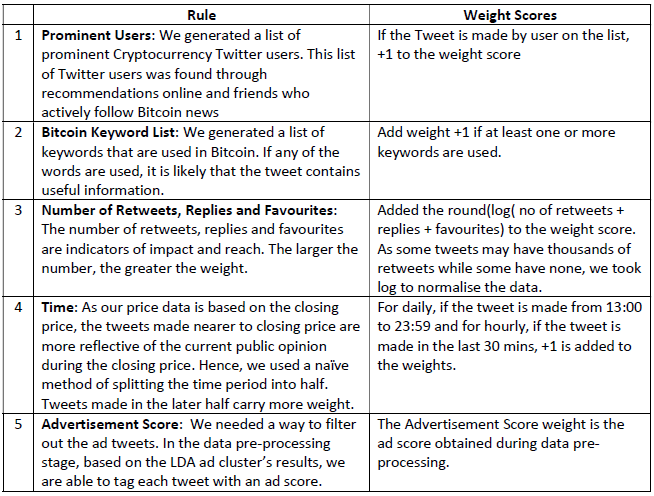}
    \label{fig:galaxy}
    \caption[Weight Rules we have established]{\textit{Weight Rules we have established}}
\end{table*}

\noindent The final weights equation is as follows:
\begin{center}
    
\textbf{Weight} = [weight(prominent user) + weight(keyword) + weight( \# of retweets) + weight( \# of   replies) + weight( \# of favorites) + weight(time)] * ad$\_$score \\
\textbf{Weighted Sentiment per tweet} = sentiment score of tweet * Weight\\
\textbf{Weighted Sentiment per day} = sum(Weighted Sentiment per tweet)/ \# of tweets in one day
\end{center}
To address the issue with VADER assigning ads high polarity scores, we multiply the ad$\_$score to the other weights. As ad$\_$score for ads are assigned as 0 and so Weighted Sentiment will be 0. As the weighted sentiment of ads are now neutral, we have effectively filtered out the ads as it will not affect our model prediction. Hence these newly generated Weighted Sentiment Scores are then passed onto the RL model as a feature.

\subsection{Reinforcement Learning}
\subsubsection{Environment}
\paragraph{Positions}
Positions of the market describe the amount of Bitcoin a particular user holds at that moment. In our environment, we have considered long and short positions. These suggest the two potential directions of the price required to maximize profit. Going Short is a trading technique in which a trader borrows an asset in order to sell it, with the expectation that the price will continue to decline. In the event that the price does decline, the short seller will then buy the asset at this lower price in order to return it to the lender of the asset, making the difference in profit. Going Long is when an investor gains exposure to cryptocurrency with the expectation that prices will rise at a later date, meaning that the asset can be sold for a profit. 

\noindent With regards to our project, the Long position wants to buy shares when prices are low and profit by sticking with them while their value is going up, and the Short position wants to sell shares with high value and use this value to buy shares at a lower value, keeping the difference as profit. 

\noindent The environment assigns a value of 0 to the agent if the position is discovered to be Short and 1 if the position is discovered to be Long. When starting a new environment, the position of the user is considered short as default. 
% \begin{figure}[H]
%     \centering
%     \includegraphics[width=7cm]{images/long_short_1.PNG}
%     \label{fig:galaxy}
%     \caption[Possible Long/Buy or Short/Sell situations]{\textit{Possible Long/Buy or Short/Sell situations}}
% \end{figure}
\smallbreak

\paragraph{Actions}
The reinforcement agent aims at maximizing profit by recommending the best course of action to the user. There are numerous actions that a user can perform in the trading market like Buy, Sell, Hold, Enter, Exit, etc. However, these actions just make things complicated with no real positive impact. In fact, they increase the learning time. Therefore there is no need to have numerous such actions, and only Sell and Buy actions are adequate to train an agent just as well. That is to say that our agent will always be recommending the user to either Buy or Sell to make sure it does not miss a single penny. The environment assigns a value of 0 to the agent if the recommended action is Sell and 1 if recommended action is Buy.

\noindent However, performing a trade on every interval does not produce reliable results for a real-life situation. To bridge the gap between simulation and reality and maximize long-term profits, the algorithm accepts the recommendation from the model and the user's current position in the market to provide the final verdict. The algorithm first decides between trading or holding. To do that, it refers to the position of the environment.

\noindent When you make a short trade, you are selling a borrowed asset hoping that its price will go down and you can buy it back later for a profit. So our algorithm will recommend the user to Buy after the short trade is satisfied and Sell after the long trade is satisfied. If the above conditions are not satisfied, then no trade occurs (Holding) because our conditional logic suggests that profit can be maximized just by holding in such a case.  

\bigbreak
\bigbreak
\bigbreak
\par\noindent\rule{8cm}{0.4pt} \\
\textbf{Algorithm 2:} Final Recommendation
\par\noindent\rule{8cm}{0.4pt} \\

\noindent \textbf{Inputs:} Recommended Action, Current Position
    
\noindent \textbf{Outputs:} Final Recommendation
\newline

    \textbf{IF} Recommended Action = Buy  \&\&  Current Position = Short \textbf{then} \\
    \indent \indent Final Recommendation = BUY \\
\newline
    \indent \textbf{ELSE IF} Recommended Action = Sell \&\&  Current Position = Short \textbf{then} \\
    \indent \indent Final Recommendation = SELL \\
\newline
    \indent \textbf{ELSE,} \\
    \indent \indent Final Recommendation = HOLD \\
\newline
    \indent \textbf{end IF} \\
\vspace{0.3cm} \\
\noindent \textbf{STOP} 
\par\noindent\rule{8cm}{0.4pt} \\

%      \hspace{2cm} \textbf{IF,}
% \begin{center}
%     Recommended Action = Buy \hspace{0.4cm} AND \hspace{0.4cm} Current Position = Short \\ OR \\ Recommended Action = Sell \hspace{0.4cm} AND \hspace{0.4cm} Current Position = Long 
% \end{center}

% \hspace{2cm} \textbf{THEN,}
% \begin{center}
%     Follow the recommend action
% \end{center}

% \hspace{2cm} \textbf{ELSE,}
% \begin{center}
%     HOLD
% \end{center}

% \begin{figure}[H]
%     \centering
%     \includegraphics[width=9cm]{images/ifcond.PNG}
%     \label{fig:galaxy}
%     \caption[Code snippet showcasing trade condition]{\textit{Code snippet showcasing trade condition}}
% \end{figure}

\noindent In the end, our algorithm can now recommend three independent actions depending on the situation of the market, namely, Buy, Hold or Sell. 

\bigbreak
\subsubsection{Learning Algorithm}

\paragraph{Model Inputs}
The reinforcement learning model predicts an action by taking the current state of the environment as input. We provide all the hand crafted features developed till now for the model to optimize its learning in Table 2.

\begin{table*}
    \centering
    \includegraphics[width=11cm]{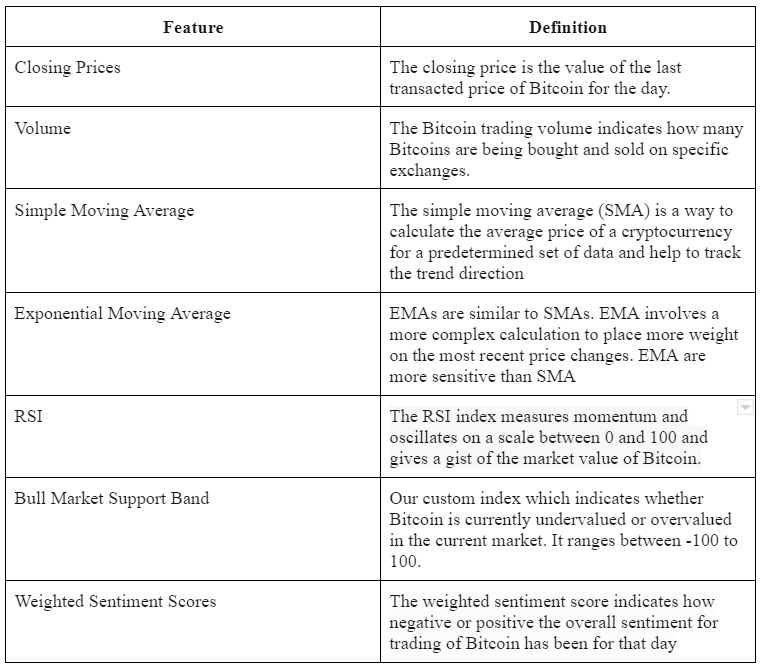}
    \label{fig:galaxy}
    \caption[Features used, along with definitions]{\textit{Features used, along with definitions}}
\end{table*}

\paragraph{Model Specifications}
The Asynchronous Advantage Actor-Critic method (A3C) has been very influential since the paper \cite{mnih2016asynchronous} was published. The algorithm combines a few key ideas:
\begin{itemize}
    \item Update rule that works on fixed-lengths of experience and uses these  to compute estimators and advantage function.
    \item Architectures that share layers between the policy and value function.
    \item Asynchronous updates.
\end{itemize}
AI researchers questioned whether the asynchrony of A3C led to improved performance or if it was just an implementation detail that allowed for faster training with a CPU-based implementation. Synchronous A2C performs better than asynchronous implementations, and we have not seen any evidence that the noise introduced by asynchrony provides any performance benefit. The host of this algorithm was the ‘stable baselines 3’ library on Python. The library offers multiple learning policies depending on the type of input data. 

\noindent We have used the MLP (Multi-Layer Perceptron) policy which acts as a base policy class for actor-critic networks allowing both policy and value prediction. It provides the learning algorithm with all the base parameters like the learning rate scheduler, activation functions, normalization, feature extraction class, and much more. This model is learned with a sliding window approach.

\section{Experiments}
\subsection{Evaluation Criteria}

\paragraph{Entropy Loss}
RL agents are known to be much more unstable to train in comparison to other types of machine learning algorithms. One of the ways that a reinforcement learning algorithm can under-perform is by becoming stuck during training on a strategy that is neither a good solution nor the absolute worst solution. This phenomenon is referred to as reaching a “local minimum”. 

\noindent In reinforcement learning, a similar problem can occur if the agent discovers a strategy with better rewards than it was receiving when it first started but far from the optimally correct strategy. This might often manifest as the agent deciding to take a single move, over and over. Entropy loss, we believe first discussed in a 1991 paper, is an additional loss parameter that can be added to a model to help with local minima in the same way that momentum might and to provide encouragement for the agent to take a variety of moves and explore the set of strategies more. 

\noindent To solve this, we will encourage the agent to vary its moves by adding a new loss parameter based on the entropy of its predicted moves. The equation for entropy here is described in Equation 5.
\begin{equation}
    H(x) = \sum_{i=1}^{n} P(x_i) \log_e P(x_i)
\end{equation}
% \begin{figure}[H]
%     \centering
%     \includegraphics[width=9cm]{images/entropy_formula.PNG}
%     \label{fig:galaxy}
%     % \caption[Entropy Formula]{}
% \end{figure}

This encourages the network to only make strong predictions if it is highly confident in them. Entropy loss is a clever and simple mechanism to encourage the agent to explore by providing a loss parameter that teaches the network to avoid very confident predictions. As the distribution of the predictions becomes more spread out, the network will sample those moves more often and learn that they can lead to greater rewards.

\bigbreak
\subsection{Training of proposed model}

The training of the RL agent was done on the Google Colab GPU, with \textit{NVIDIA-SMI 460.67       Driver Version: 460.32.03    CUDA Version: 11.2.} The model was tested for multiple scenarios, and the below gave the most consistent results. The time steps were set to be 50000, which is equivalent to the number of iterations of training. The 1918 data instances were present, and 1233 were used for training the model with a window size of 30. The remaining 685 instances were used for testing. 
% 
% \subsection{User Interface}

% A web app is deployed to allow the RL agent to recommend the optimal action to the user. The agent can recommend one of the following actions depending on the market situation: Buy, Hold, Sell. 
% \begin{figure}[H]
%     \centering
%     \includegraphics[width=10cm]{images/web_recomd (1).PNG} 
%     \label{fig:galaxy}
%     \caption[Website Recommendation]{}
% \end{figure}

% \noindent The website also provides visualizations of various important indicators for the user. This would further assist the user to reinforce the agent's recommendation.
% \begin{figure}[H]
%     \centering
%     \includegraphics[width=10cm]{images/close_price.png}
%     \label{fig:galaxy}
%     \caption[Closing Prices on website]{}
% \end{figure}

% \begin{figure}[H]
%     \centering
%     \includegraphics[width=10cm]{images/rsi.png}
%     \label{fig:galaxy}
%     \caption[RSI on website]{}
% \end{figure}

% \begin{figure}[H]
%     \centering
%     \includegraphics[width=10cm]{images/sma.png}
%     \label{fig:galaxy}
%     \caption[SMA on website]{}
% \end{figure}

% \begin{figure}[H]
%     \centering
%     \includegraphics[width=10cm]{images/ema_new.png}
%     \label{fig:galaxy}
%     \caption[EMA on website]{}
% \end{figure}

% \begin{figure}[H]
%     \centering
%     \includegraphics[width=10cm]{images/sentime.png}
%     \label{fig:galaxy}
%     \caption[Sentiment scores on website]{}
% \end{figure}

% \begin{figure}[H]
%     \centering
%     \includegraphics[width=10cm]{images/newplot.png}
%     \label{fig: BMSB}
%     \caption[BMSB Gauge on website]{}
% \end{figure}

\bigbreak

\subsection{Results and discussion}

After training for 50000 time steps, the entropy loss of 0.0209 was obtained. The learning rate was optimized at 0.007.

% \begin{figure}[H]
%     \centering
%     \includegraphics[width=5cm]{images/loss_metrix.PNG}
%     \label{fig:galaxy}
%     \caption{Loss for training}
% \end{figure}

Finally, the testing was performed on the latest 685 days in the data. Specifically, the testing period started from 17th May 2019 and ended on 1st April 2021. This period is highly significant as it captures the peak points of Covid-19 and its effects on the market. Given these extreme scenarios, the model was able to generate a total reward of 39,412 and a profit value of 1.69486, which is equivalent to a 69.486\% increase over 685 days. \\
\begin{figure*}
    \centering
    \includegraphics[width=\textwidth,height=7cm]{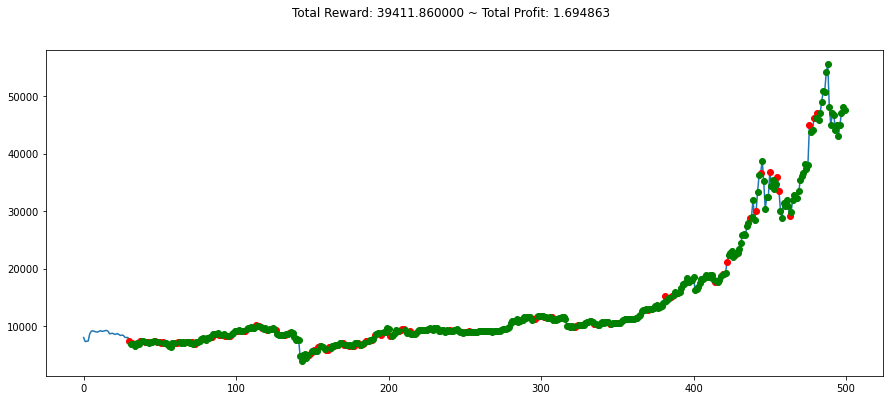}
    \label{fig:galaxy}
    \caption{Graph of recommendations}
\end{figure*}
The green dots in the Fig. 4 indicate a long position, whereas the red dots indicate a short position. As discussed before, the agent itself recommends only Buy or Sell with the aim to capture every penny. Our algorithm then reviews the current situation/position of the user in the market and either recommend the agent action or asks them to hold their assets.

\section{Conclusion}

% \subsection{Conclusion}
Investing in the crypto market is a tedious task and requires a lot of effort to make good trading strategies. Moreover, analyzing real-time data is a more difficult task. Traders make several trading decisions and keep on learning from them. During this process, they improve their decision-making skills. This is the major idea behind implementing a Reinforcement Learning agent, which trains against the environment to adapt and make better decisions. Thus, we can say that trading can be approached as a Reinforcement Learning Problem.

\noindent In the trading scene, experts use sophisticated tools to analyze the trend of prices better. Our plan was to allow the learning agent to learn in such a feature-rich learning environment. Specifically, we have used some of the most popular and tested long-term trend indicators like SMA, RSI, and EMA. We have also crafted our technical indicator, which is used by the agent to find opportunities for trade action. Another huge factor that impacts the flow of prices is the overall sentiment of the market. By including this factor in the environment, the agent was better able to understand the market situation. 

\noindent The value of the added features was demonstrated as without these, the average profit of the model was not always positive. In addition to the features, we were able to get consistently positive profits over a long period of testing. With these handcrafted features, our model was able to analyze the situation of the user and market and recommend smart decisions to the user.

\noindent We were able to provide an interface to the user for their better understanding of the current market situation through visualizations of the important indicators and sentiment scores. The interface was able to host the agent to provide its final recommendation to the user. The results show great potential for the approach, but the bitcoin markets are quite large, complex and volatile, so the modelling of this environment still presents a lot of challenges. 

\section{Further work}
Many experts in the field of cryptocurrency and stock trading utilize trend analysis by identifying the popular patterns in the price action. Each of these patterns helps in the analysis of the price changes to occur in the future. The ascending and descending triangle pattern, as shown in the figure, leads to a continuity in the trend of prices. Another popular pattern experts use is the head and shoulders pattern shown in figure 2. This pattern is a strong sign of a reversal of the trend of the prices. Because of their impact, the recognition of these patterns becomes of most significance. A reliable recognition system will be sure to aid the reinforcement agent in better understanding the trading space and making smarter decisions. However, current pattern matching algorithms fail to work for different pattern spans. This problem is highly significant as even though we have an idea of what pattern we are looking for, most patterns occur at significantly different intervals. 

\noindent Nevertheless, there is still a promise of research in this department to take the project to the next level. In recent times, communities on Reddit have had a significant impact on the prices of cryptocurrencies. So another step forward for the project would be to include sentiments from specific subReddits to our sentiment scores. This will involve assigning weights to communities and comparing them with the scores from Twitter.
\section*{Acknowledgments}

We would like to express my very great appreciation to Ankit Khivasara Sir for their valuable and constructive suggestions during the planning and development of this project. Their willingness to give their time so generously has been very much appreciated.

\noindent We would also like to extend our gratitude to the entire team of OpenAI as well as that of Stable Baselines3 for implementing and maintaining their framework, without which this project would not be possible.

% \section*{References}

% Please number citations consecutively within brackets \cite{b1}. The 
% sentence punctuation follows the bracket \cite{b2}. Refer simply to the reference 
% number, as in \cite{b3}---do not use ``Ref. \cite{b3}'' or ``reference \cite{b3}'' except at 
% the beginning of a sentence: ``Reference \cite{b3} was the first $\ldots$''

% Number footnotes separately in superscripts. Place the actual footnote at 
% the bottom of the column in which it was cited. Do not put footnotes in the 
% abstract or reference list. Use letters for table footnotes.

% Unless there are six authors or more give all authors' names; do not use 
% ``et al.''. Papers that have not been published, even if they have been 
% submitted for publication, should be cited as ``unpublished'' \cite{b4}. Papers 
% that have been accepted for publication should be cited as ``in press'' \cite{b5}. 
% Capitalize only the first word in a paper title, except for proper nouns and 
% element symbols.

% For papers published in translation journals, please give the English 
% citation first, followed by the original foreign-language citation \cite{b6}.

% \section*{References}
% \newpage
% \section*{Bibliography}
\addcontentsline{toc}{section}{References}
\printbibliography 

% \bibitem{b1} G. Eason, B. Noble, and I. N. Sneddon, ``On certain integrals of Lipschitz-Hankel type involving products of Bessel functions,'' Phil. Trans. Roy. Soc. London, vol. A247, pp. 529--551, April 1955.
% \bibitem{b2} J. Clerk Maxwell, A Treatise on Electricity and Magnetism, 3rd ed., vol. 2. Oxford: Clarendon, 1892, pp.68--73.
% \bibitem{b3} I. S. Jacobs and C. P. Bean, ``Fine particles, thin films and exchange anisotropy,'' in Magnetism, vol. III, G. T. Rado and H. Suhl, Eds. New York: Academic, 1963, pp. 271--350.
% \bibitem{b4} K. Elissa, ``Title of paper if known,'' unpublished.
% \bibitem{b5} R. Nicole, ``Title of paper with only first word capitalized,'' J. Name Stand. Abbrev., in press.
% \bibitem{b6} Y. Yorozu, M. Hirano, K. Oka, and Y. Tagawa, ``Electron spectroscopy studies on magneto-optical media and plastic substrate interface,'' IEEE Transl. J. Magn. Japan, vol. 2, pp. 740--741, August 1987 [Digests 9th Annual Conf. Magnetics Japan, p. 301, 1982].
% \bibitem{b7} M. Young, The Technical Writer's Handbook. Mill Valley, CA: University Science, 1989.
% \end{thebibliography}

\end{document}